\documentclass[12pt]{article}
\usepackage{epsf}
\usepackage{cite}

\newcommand{\bmat}{\left(\begin{array}}
\newcommand{\emat}{\end{array}\right)}

\def\yzero{\smash{\hbox{$y\kern-4pt\raise1pt\hbox{${}^\circ$}$}}}

\def\beq{\begin{equation}}
\def\eeq{\end{equation}}
\def\beqa{\begin{eqnarray}}
\def\eeqa{\end{eqnarray}}

\def\-{\hphantom{-}}

\def\s2{\frac{1}{\sqrt2}}

\def\beq{\begin{equation}}
\def\eeq{\end{equation}}
\def\beqa{\begin{eqnarray}}
\def\eeqa{\end{eqnarray}}

\def\IF{\relax{\rm I\kern-.18em F}}
\def\II{\relax{\rm I\kern-.18em I}}
\def\IP{\relax{\rm I\kern-.18em P}}
\def\IC{\relax\hbox{\kern.25em$\inbar\kern-.3em{\rm C}$}}
\def\IR{\relax{\rm I\kern-.18em R}}

\def\Dsl{\,\raise.15ex\hbox{/}\mkern-13.5mu D} 
\def\IZ{Z\kern-.4em  Z}

\catcode`\@=11
\newdimen\@rotdimen
\newbox\@rotbox

\def\@vspec#1{\special{ps:#1}}
\def\@rotstart#1{\@vspec{gsave currentpoint currentpoint translate
   #1 neg exch neg exch translate}}
\def\@rotfinish{\@vspec{currentpoint grestore moveto}}
\def\@rotr#1{\@rotdimen=\ht#1\advance\@rotdimen by\dp#1%
   \hbox to\@rotdimen{\hskip\ht#1\vbox to\wd#1{\@rotstart{90 rotate}%
   \box#1\vss}\hss}\@rotfinish}
\def\@rotl#1{\@rotdimen=\ht#1\advance\@rotdimen by\dp#1%
   \hbox to\@rotdimen{\vbox to\wd#1{\vskip\wd#1\@rotstart{270 rotate}%
   \box#1\vss}\hss}\@rotfinish}%
\def\@rotu#1{\@rotdimen=\ht#1\advance\@rotdimen by\dp#1%
   \hbox to\wd#1{\hskip\wd#1\vbox to\@rotdimen{\vskip\@rotdimen
   \@rotstart{-1 dup scale}\box#1\vss}\hss}\@rotfinish}%
\def\@rotf#1{\hbox to\wd#1{\hskip\wd#1\@rotstart{-1 1 scale}%
   \box#1\hss}\@rotfinish}%
\def\rotate{\@ifnextchar[{\@rotate}{\@rotate[l]}}
\def\@rotate[#1]#2{\setbox\@rotbox=\hbox{#2}\@nameuse{@rot#1}\@rotbox}

\catcode`\@=12

\topmargin
-1.5cm
\textwidth
15.5cm
\textheight
23.5cm
\oddsidemargin
0.7cm
\evensidemargin
1.2cm

\begin{document}

\makeatletter
\@addtoreset{equation}{section}
\makeatother
\renewcommand{\theequation}{\arabic{equation}}
\pagestyle{empty}
\rightline{FTUAM-02/29; IFT-UAM/CSIC-02-48}
\rightline{\tt hep-ph/0212048}
\vspace{0.5cm}
\begin{center}
\LARGE{ More about the Standard Model at Intersecting Branes 
\footnote{Contribution to the proceedings of SUSY-02, Hamburg.
}
\\[10mm]}
\large{ 
D. Cremades, L.E. Ib\'a\~nez and F. Marchesano
\\[2mm]}
\small{
 Departamento de F\'{\i}sica Te\'orica C-XI
and Instituto de F\'{\i}sica Te\'orica  C-XVI,\\[-0.3em]
Universidad Aut\'onoma de Madrid,
Cantoblanco, 28049 Madrid, Spain.
\\[9mm]}
\small{\bf Abstract} \\[7mm]
\end{center}

\begin{center}
\begin{minipage}[h]{14.0cm}
Intersecting D-brane models seem to be one of the
most promising avenues to embed the Standard Model
physics within the string context.
We review here different aspects of these models.
Topics include the question of SUSY and quasi-SUSY
in intersecting brane models, model-building,
the brane recombination interpretation of the SM Higgs 
mechanism, Yukawa couplings, the lowering of the string scale  
and possible new Z's accessible to accelerators.

\end{minipage}
\end{center}
\newpage
\setcounter{page}{1}
\pagestyle{plain}
\renewcommand{\thefootnote}{\arabic{footnote}}
\setcounter{footnote}{0}



In the last couple of years there have been renovated efforts 
in looking  for  D-brane
configurations with a low-energy
effective theory resembling the standard model (SM).
One approach which looks particularly successful is 
that of intersecting D-brane models \cite{bgkl,afiru,afiru2,bkl,imr,dubna}
(see also \cite{bklo,csu,bailin,cim1,hon,raul,cim2,koko,cim3,ekn}). 
We will not attempt to give here 
an introduction to the subject (see  e.g.  \cite{dubna}
and references therein).
Rather we will concentrate on giving a brief report on some 
of the work in the subject that we have been involved with
in the last year.

\begin{figure}
\centering
\epsfxsize=3.4in
\hspace*{0in}\vspace*{.2in}
\epsffile{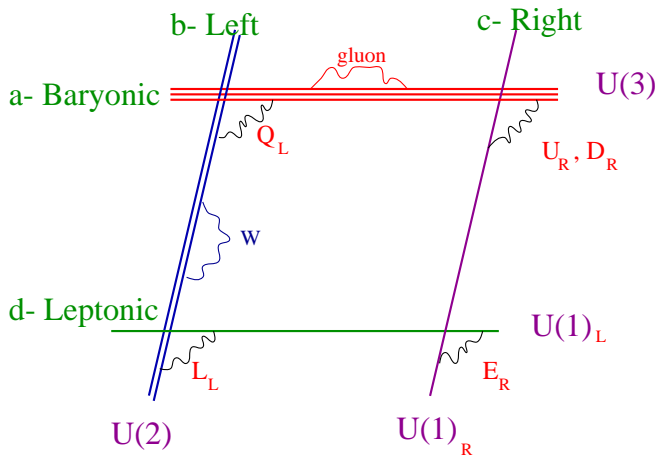}
\caption{\small
The SM spectrum at intersecting branes.}
\label{2}
\end{figure}

In intersecting brane models (fig. \ref{2}) the different gauge 
interactions live 
on different stacks of  D-branes, the simplest configurations 
having four stacks: {\it baryonic, left, right and leptonic}.
In particular  one considers  stacks of branes with multiplicities
$N_a=3$, $N_b=2$, $N_c=1$, $N_d=1$, yielding initially a  gauge group 
$U(3)\times U(2)\times U(1)\times U(1)$. Up to three of the $U(1)$'s
become massive by combining with some closed string (Ramond-Ramond)
fields so that in the simplest situation one is just left with
standard hypercharge and the full group is that of the SM.
The D$p$-branes (with $p$ = 6, 5, 4) worldvolumes contain  Minkowski
space and the remaining $(p-3)$ dimensions wrap cycles on compact
dimensions.  At the brane intersections in extra dimensions
live quarks and leptons and the triplication of generations 
appear because in the compact space 
the different branes intersect three times.
In particular, if we denote by $I_{ab}$ the number of times 
that branes $a$ and $b$ intersect, the following 
intersection numbers \cite{imr}:
\beq
\begin{array}{lll}
I_{ab} = 1, & & I_{ab*} = 2, \\
I_{ac} = -3, & & I_{ac*} = -3,  \\
I_{bd} = 0, & & I_{bd*} = - 3, \\
I_{cd} = -3 & & I_{cd*} = 3,
\end{array}
\label{intersec2}
\eeq
give rise to the fermion spectrum of the SM.
Specific D6-brane models in which the compact space is just a 
6-torus $T^6$ and yielding the above SM spectrum were
provided in ref.\cite{imr}. One can also find D5-brane models in which
the compact space  is $T^4 \times (T^2/{\bf Z}_N)$ and one 
obtains the same SM chiral fermion spectrum \cite{cim3}.
In both classes of constructions there is an interesting 
connection between the number of generations and the number
of colours. Indeed, in order to cancel anomalies the net number
of $U(2)_b$ doublets has to equal that of anti-doublets \cite{imr},
which in these models happens only because the number of generations 
equals the number of colours. In addition, 
one of the nicest features of these constructions is that 
the proton is automatically stable since baryon number 
($U(1)_a$) is a gauged symmetry \cite{imr}.

\begin{figure}
\centering
\epsfxsize=3.0in
\hspace*{0in}\vspace*{.2in}
\epsffile{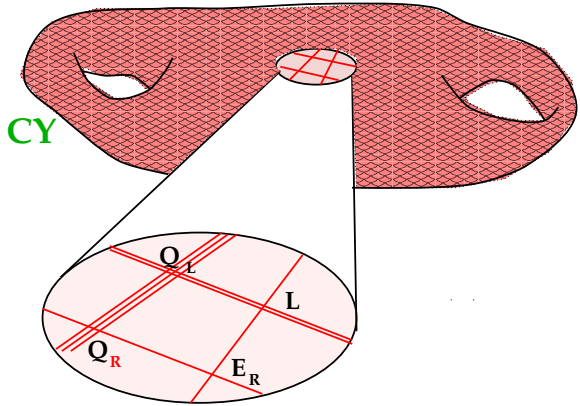} 
\caption{\small
The general intersecting brane system may be embedded, e.g,
into a general Calabi-Yau compact space.}   
\label{cy}
\end{figure}

We must emphasize  that the intersecting brane
 setting just described is quite general. 
As fig. \ref{cy} illustrates,
one may consider for example  four stacks $a, b, c, d$ 
of D6-branes   wrapping 3-cycles on a  
complicated Calabi-Yau. As long as the intersection numbers 
are as above,  the chiral fermion spectrum is going to be
the one of the SM  independently  of the details of the compactification.
As an example the above SM spectrum has also been recently obtained 
from D6-branes wrapping 3-cycles on the quintic CY \cite{bbkl} or 
in certain non-compact manifolds \cite{uranga}.

The mentioned specific toroidal models are generically non-SUSY 
 due to the presence of
the  intersections. Thus, for example, associated to each    
of the intersections there are  massive scalar fields
which in some sense may be considered  "SUSY-partners", squarks and sleptons,
of the massless chiral fermions, have the same multiplicity $|I_{ij}|$
 and carry the same gauge quantum numbers.
The lightest of those states have  masses
\cite{afiru}
{\small \beqa
\begin{array}{cc}
 &  {\rm (Mass)}_1^2 =
\frac 12(-\vartheta_1+\vartheta_2+\vartheta_3)M_s^2 \\
 &  {\rm (Mass)}_2^2 =
\frac 12(\vartheta_1-\vartheta_2+\vartheta_3)M_s^2 \\
  &  {\rm (Mass)}_3^2 =
\frac 12(\vartheta_1+\vartheta_2-\vartheta_3)M_s^2
\label{tachdsix}
\end{array}
\eeqa}
where $\vartheta_i $ are the abslute  value of the 
intersection angles (in units of $\pi $) at each of the three
subtori. 
As is obvious
from these formulae  the masses depend on the angles at each intersection
and hence on the relative size of the radii.
In  principle some of the scalars could  be tachyonic,
 but  in general it is possible
   to vary the compact radii in order  to get rid of all tachyons
of a given model (see \cite{afiru,imr}).
On the other hand, one can also adjust the radii so that there is one 
massless scalar at the intersection. Then one gets $N=1$ SUSY
{\it at that specific intersection}. Is it possible to get
a fully $N=1$ SUSY model, i.e.,  
a model  in which all intersections respect the {\it same } 
$N=1$ supersymmetry? The answer is  no, at least in the purely
toroidal examples as in ref.\cite{imr} . The reason for this is 
that D-brane configurations wrapping compact spaces as in here
have to respect the conditions of RR-tadpole cancellation.
The overall charge of the configuration with respect to certain
tensorial RR-fields has to vanish. In the purely toroidal case those 
conditions turn out to be incompatible with the geometrical
configurations required to get $N=1$ SUSY.
Fully $N=1$ SUSY intersecting brane models may be built in the case  with 
an added ${\bf Z}_2\times {\bf Z}_2$ orbifold twist \cite{csu}. 
However, in that case 
additional chiral  exotics beyond the SM content seem unavoidable.

\begin{figure}
\centering
\epsfxsize=3.0in
\hspace*{0in}\vspace*{0in}
\epsffile{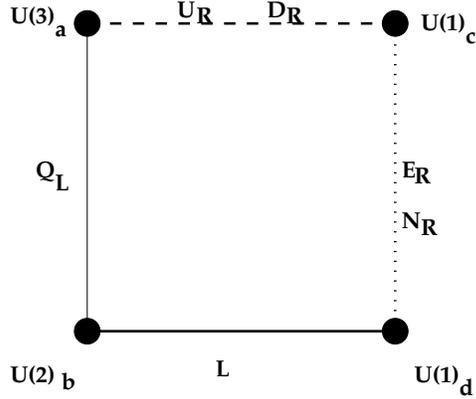}
\caption{\small A square SUSY-quiver. }
\label{qsimr}
\end{figure}

On the other hand there is an interesting possibility termed
quasi-SUSY in ref.\cite{cim1}  and pseudo-SUSY in ref.\cite{klein}. The
possibility exists 
that all intersections respect some $N=1$ SUSY but {\it different ones}.
In the case of toroidal models this possibility is still compatible 
with cancellation of RR tadpoles. For example, certain
subset of the models in ref.\cite{imr} can be made quasi-SUSY by choosing 
appropriate radii. Let us 
 denote by $(n_a^i,m_a^i)$, $i$ = 1, 2, 3
the wrapping numbers of each brane $D6_a$, $n_a^i$($m_a^i$) being the number
of times the brane is wrapping around the  $x$($y$)-coordinate of the $i-th$
torus. Consider in particular the  wrapping numbers for the different branes
given in table \ref{QSUSY}.
\begin{table}[htb] \footnotesize
\renewcommand{\arraystretch}{2}
\begin{center}
\begin{tabular}{|c||c|c|c|}
\hline
 $N_i$    &  $(n_i^1,m_i^1)$  &  $(n_i^2,m_i^2)$   & $(n_i^3,m_i^3)$ \\
\hline\hline $N_a=3$ & $(1,0)$  &  $(n_a^2, \beta^2)$ &
 $(3 ,  -1/2)$  \\
\hline $N_b=2$ &   $(n_b^1, 1)$    &  $ (1/\beta^2,0)$  &
$(1,-1/2)$   \\
\hline $N_c=1$ & $(0,1)$  &
 $(1/\beta^2,0)$  & $(0,1)$  \\
\hline $N_d=1$ &   $(1,0)$    &  $(n_a^2,3\beta^2 )$  &
$(1, 1 /2)$   \\
\hline $N_h$ &   $(1,0)$    &  $(1/\beta^2,0 )$  &
$(n_h^3,m_h^3)$   \\ 
\hline \end{tabular}
\end{center} \caption{\small D6-brane wrapping numbers giving rise to a Q-SUSY
SM spectrum for a square quiver. For the sake of generality we have
also considered the possible presence of an extra brane
with no intersection with the SM branes.
\label{QSUSY} }
\end{table}
The number of times the two branes
$D6_a$  and $D6_b$ intersect in $T^6$ is given by the intersection number
\cite{bgkl} $
I_{ab}\ =\
(n_a^1m_b^1-m_a^1n_b^1)$$(n_a^2m_b^2-m_a^2n_b^2)$$(n_a^3m_b^3-m_a^3n_b^3)$.
Then one can easily check that this brane setting yields the 
chiral spectrum of the SM. Now, the masses of the scalars depend on the ratios
$U^i=R_2^i/R_1^i$, $i=1,2,3$. One can easily check that if we
set $U^1=(n_b^1/2)U^3$ and $U^2=(n_a^2/6\beta^2)U^3$  massless scalars
appear at each intersection and some $N=1$ SUSY is preserved at each of them.
This property may be depicted in terms of a {\it square quiver diagram}, 
shown in fig. \ref{qsimr}.
Thus, now each quark and lepton has a massless SUSY-partner, 
very much as in the SUSY-SM. The model is however not
fully $N=1$ supersymmetric because each intersection respects
a different SUSY. This kind of {\it quasi-SUSY} theories have 
some interesting properties. In particular, loop corrections 
to scalar masses appear only at two loops, since only at that order
the global non-SUSY structure of the configuration may be 
noticed \cite{cim1,cim2,klein}. This loop suppression of scalar masses may 
be interesting phenomenologically in order to address the 
``modest hierarchy problem'', i.e., 
in order  to maintain a hierarchy between  a string
scale of order 10-100 TeV and the electroweak scale.

There is a variety of SUSY-quivers that one may consider leading to
different low-energy models \cite{cim2}. One can also  find D6-brane
configurations
wrapping $T^6$ leading to the massless chiral spectrum of the MSSM.
Some examples were presented in ref.\cite{cim1,cim2} but we 
will present here a new and remarkably  simple model which will be 
discussed in more detail elsewhere \cite{cim4}.
Consider the stacks of D6-branes with the wrapping numbers of 
table \ref{QSUSYsquare}.
\begin{table}[htb] \footnotesize
\renewcommand{\arraystretch}{1.5}
\begin{center}
\begin{tabular}{|c||c|c|c|}
\hline
 $N_i$    &  $(n_i^1,m_i^1)$  &  $(n_i^2,m_i^2)$   & $(n_i^3,m_i^3)$ \\
\hline\hline $N_a=3$ & $(1,0)$  &  $(1/\rho , 3\rho )$ &
 $(1/\rho  ,  -3\rho )$  \\
\hline $N_b=1$ &   $(0, 1)$    &  $ (1,0)$  & 
$(0,-1)$   \\
\hline $N_c=1$ & $(0,1)$  & 
 $(0,-1)$  & $(1,0)$  \\
\hline $N_d=1$ &   $(1,0)$    &  $(1/\rho ,3\rho )$  &
$(1/\rho , -3\rho )$   \\
\hline \end{tabular}
\end{center} \caption{\small D6-brane wrapping numbers giving rise to a
the chiral spectrum of the MSSM. Here $\rho = 1,1/3$.
\label{QSUSYsquare} }
\end{table}
Generically the gauge group of this configuration is 
$U(3)\times U(1)^3$. However, one can check that the 
symmetry is enhanced to $U(3)_a\times SU(2)_b\times U(1)_c\times U(1)_d$
if the brane $b$ is located on top of its orientifold mirror
$b^*$. Computing the intersection numbers as above one gets
the result
\beq
\begin{array}{lll}
I_{ab} = 3, & & I_{ab*} = 3, \\
I_{ac} = -3, & & I_{ac*} = -3,  \\
I_{db} = 3, & & I_{db*} = 3, \\
I_{dc} = -3 & & I_{dc*} = 3,\\
I_{bc} = -1 & & I_{bc*} = 1,
\end{array}
\label{intersecMSSM}
\eeq
which corresponds to the 
 chiral fermion spectrum of the SM (plus right-handed neutrinos).
In addition there is a minimal set of Higgs multiplets if 
one locates the brane $b$ on top of the brane $c$ along the first
torus. In other words, there is a minimal Higgs sector with a 
$\mu$-parameter given by the distance between branes 
$b$ and $c$ along the first torus. If the ratios 
of radii in the second and third torus are equal ($U^2=U^3=\chi$)
one can check that {\it the same N=1 SUSY} is preserved at all
intersections. So this configuration is (locally) $N=1$ 
supersymmetric, and {\it the massless chiral spectrum is that 
of the MSSM with a minimal Higgs set.}
The spectrum is anomaly-free in the sense that
there are as many fundamentals as antifundamentals of any
of the groups. On the other hand the configuration cannot be made 
fully $N=1$ supersymmetric, because it turns out that 
in order to cancel RR-tadpoles an additional massive $N=0$
sector has to be added (see ref.\cite{cim2} for a discussion of 
this point). In this model there are three $U(1)$'s and only 
one of them $(3B+L)$ is anomalous and gets massive 
by combining with one RR-field. There are two massless $U(1)$'s
corresponding to $(B-L)$ and the 3-d component of right-handed
weak isospin ($U(1)_c$). So the actual low-energy gauge group
is $SU(3)\times SU(2)\times U(1)_{B-L}\times U(1)_c$.

One of the nice features of the intersecting brane approach is that 
the low-energy Lagrangian parameters admit a simple 
geometrical interpretation. We already saw an example: the 
$\mu $-parameter in this model corresponds to the distance 
between branes $b$ and $c$ in the first torus.
 Another example is
the generation of tree level Fayet-Iliopoulos terms
for the anomalous $U(1)$'s \cite{csu,cim1}.
If one has a small departure from the 
SUSY geometry, i.e., if $U^2=U^3+\delta $ with $\delta $ small,
one finds a FI-term for the 
anomalous U(1) \cite{cim1} :
\beq
\xi \ =\ M_s^2 \times \frac {3\rho^2 \delta }
{1+(3\rho^2U^3)^2 } \ .
\eeq
If $\delta \not= 0$, the existence of this FI-term may trigger
further gauge symmetry breaking. In particular the additional 
$U(1)$ may be broken down to standard hypercharge by 
inducing a vev to the right-handed sneutrino.

\begin{figure}
\begin{center}
\centering
\epsfysize=2.3in
\leavevmode
\epsfbox{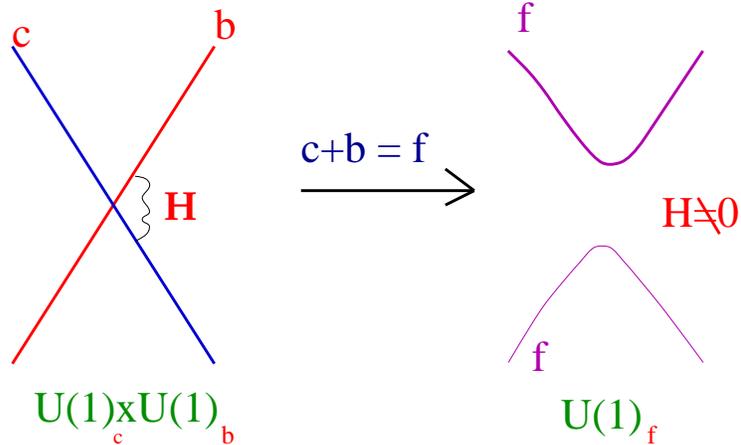}
\end{center}
\caption{\small
Branes $b$ and $c$ are recombined into a single brane $f$.
The gauge symmetry is reduced.}
 \label{recomb}
\end{figure}

Before taking into account SUSY-breaking effects, the
above local SUSY configuration has 
(for vanishing $\mu$-term) a flat direction in which
electroweak symmetry is broken by the vevs of the Higgs scalars
lying at the $bc$ and $bc$* intersections
$H_{bc}$.  The Higgs mechanism 
has also a geometric interpretation in terms of branes \cite{cim2}.
As illustrated in fig. \ref{recomb}, 
a vev for the Higgs fields $H_{bc}$ corresponds to a process 
$b+c\rightarrow f$ in which branes $b$ and $c$ recombine into
a single brane $f$. Since we have started from two branes
(plus orientifold mirrors) and end up with one brane and its mirror,
the rank of the gauge group has been reduced. Altogether we are left at the
end of the day with only three brane stacks, a,d,f and 
a gauge group $SU(3)\times U(1)_{em}$ at the massless level.

\begin{figure}
\begin{center}
\centering   
\leavevmode
\epsfxsize= 4.5in
\epsfbox{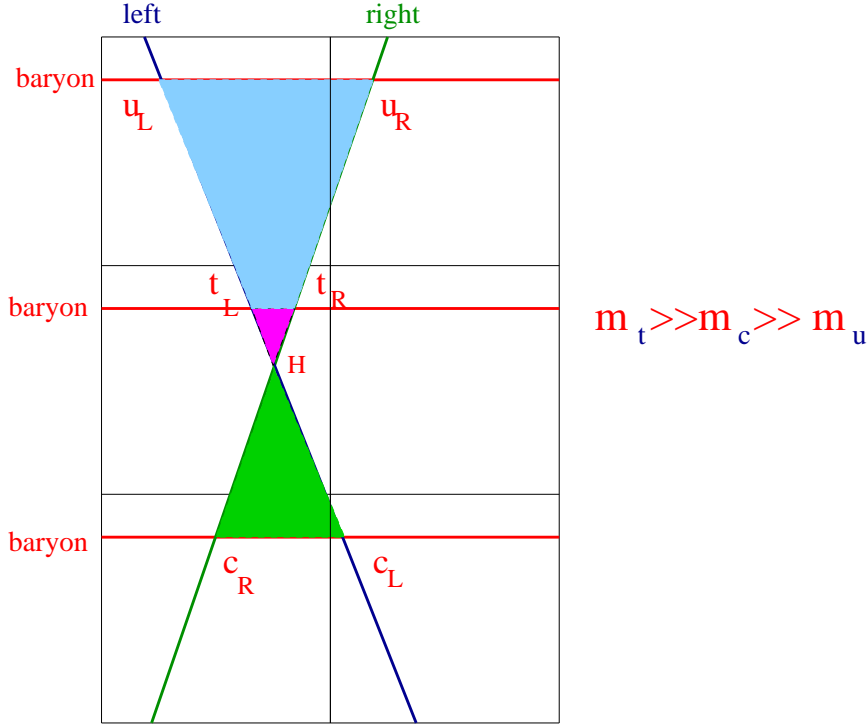}
\end{center}
\caption{\small  The Yukawa couplings are computed from correlators
involving a Higgs field, a right-handed fermion and
a left-handed fermion. Open strings have to stretch 
in worldsheets of triangle shape. Those triangles have
different size for different generations, leading to 
Yukawa textures and hierarchies (see ref.\cite{cim4}).}
 \label{yuk}
\end{figure}

The Yukawa couplings also have an interesting geometrical
interpretation in intersecting brane models \cite{afiru2}.
 A Yukawa coupling 
involves correlators of a Higgs field, a right-handed fermion 
and a left-handed fermion. The worldsheet of strings 
connecting those three vertices has a triangular shape,
as in fig. \ref{yuk}, with open strings stretching 
between the three intersecting branes participating in 
the coupling. 
 The Yukawa couplings are then  
proportional to $exp(-S_{cl})$, $S_{cl}$ being the classical 
string action, which is proportional to the area of the 
worldsheet. This provides a nice physical origin for the
observed hierarchy of fermion masses  since,
as exemplified in fig. \ref{yuk}, 
 the size of the relevant 
triangles for the different generations is in general different
\cite{afiru2}.
Thus, e.g., the triangle associated to the top-quark
coupling would be smaller than the one associated to the 
c-quark which would, in turn, be smaller than the one 
of the u-quark.
One can also see that generation mixing as well as 
complex phases do in general appear.
An analysis of Yukawa couplings in intersecting
brane models will appear in ref.\cite{cim4}.
%

\begin{figure}
\centering
\epsfxsize=4.9in
\hspace*{0in}\vspace*{.2in}
\epsffile{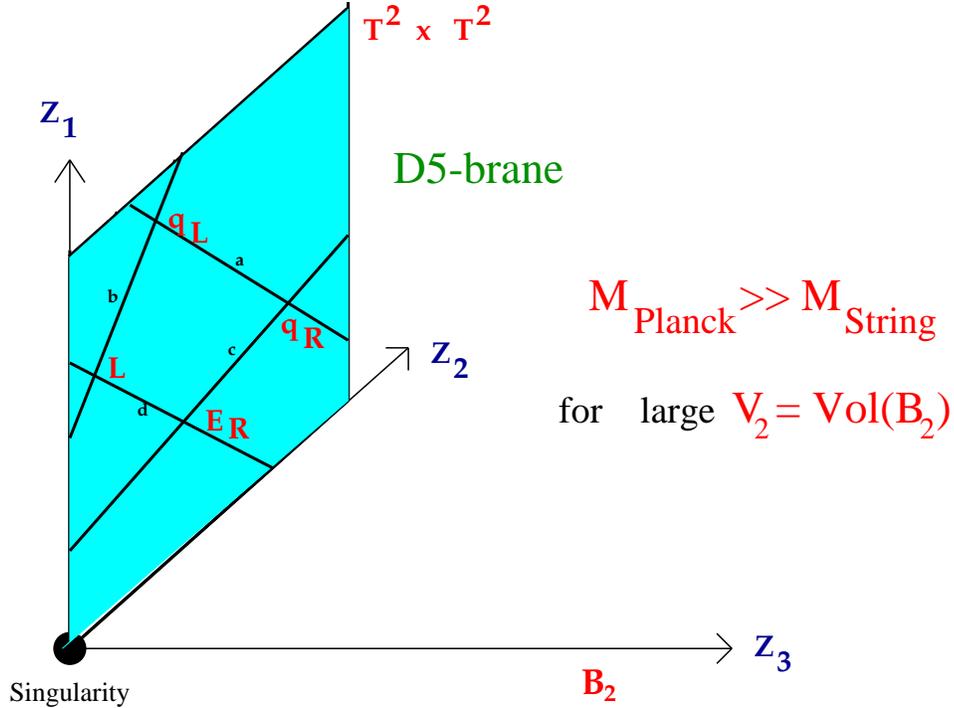}
\caption{\small Intersecting D5-world set up.
One can obtain a low string scale $M_s<<M_p$ if
the volume of the two transverse dimensions is large.}
\label{braneworld}
\end{figure}

One interesting question is whether in this class of
intersecting brane models  one can realize the 
low string scale scenario \cite{aadd} with $M_s\sim 1-10$ TeV.
This is particularly relevant in models which are 
not supersymmetric and in which lowering the string scale
down to the TeV scale provides then a solution to the
hierarchy problem.
As is well known, this requires that at least  2 
of the 6 compact dimensions become very large,
so that we obtain a large splitting between the string scale
$M_s$ and the effective 4-dimensional gravity scale 
$M_{Planck}$. In the simple  case of D6-branes wrapping
a 6-torus, realizing the low string scale scenario is
in principle complicated \cite{bgkl}. This is because there are no
 dimensions which can be made large and are orthogonal to the  
SM brane system. On the other hand, as pointed out in 
\cite{afiru} and explained in more detail in
\cite{uranga}, one can start from a toroidal model as above 
and construct  a related model in which the compact 
volume can be made arbitrarily large without affecting the
SM branes. This is done by glueing an infinite  throat 
to the torus in a region far away from the branes 
and then connecting the throat to some large volume.
Alternatively one can consider other intersecting brane
constructions in which the presence of a couple of  dimensions
transverse to the SM branes which can be made large is more obvious. 
For example, one can consider Type IIB compactified on
$T^2\times T^2\times (T^2/{\bf Z}_N)$ with D5-branes wrapping 2-cycles on 
$T^2\times T^2$ and located at a fixed point of the orbifold 
$T^2/{\bf Z}_N$ (see fig. \ref{braneworld}). 
One can  now have a low string scale $M_s\propto 1-10$ TeV
while maintaining the experimentally measured four-dimensional
Planck mass $M_p= 1.18 \times 10^{19}$ GeV by
some dimensions getting very large \cite{aadd}.
 Let us denote by $V_4$ the volume of $T^4$ and by
$V_2$ that of $T^2/{\bf Z}_N$. Then the Planck scale
is given by
\beq
M_p \ =\  {2\over \lambda }  M_s^4 \sqrt{V_4V_2}
\eeq
In order to avoid too light KK/Winding modes in the   
worldvolume of the D5-branes let us assume $V_4\propto 1/M_s^4$.
Then one has
\beq
V_2 \ =\  {  { M_p^2 \lambda ^2 } \over {4 M_s^4} }
\eeq
and one can accommodate a low string scale $M_s\propto 1$ TeV
by having the volume $V_2$  of the 2-dimensional manifold ${\bf B_2}$
large enough (i.e., of order $(mm.)^2$). 
Such D5-brane models with the chiral fermion content 
as in eq.(\ref{intersec2}) leading to the fermions of the SM
have been recently constructed in ref.\cite{cim3}.
Note, however, that those D5-brane constructions are intrinsically
non-SUSY since  the ${\bf Z}_N$ orbifold projection projects out all
gauginos.

\begin{figure}
\centering
\epsfxsize=4in
\hspace*{0in}\vspace*{.2in}
\epsffile{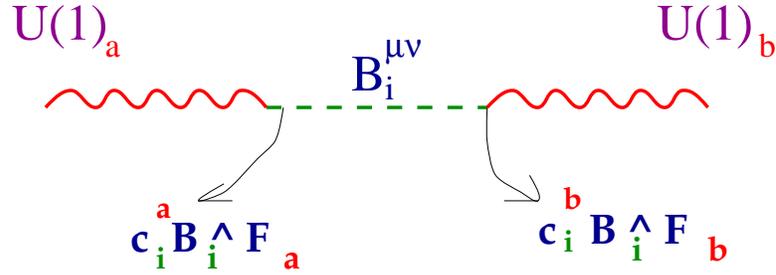}
\caption{\small  Mixing of $U(1)$'s and antisymmetric RR-fields 
$B_i^{\mu \nu }$ yielding masses to Abelian gauge fields.}
\label{stuck}
\end{figure}

In the case of a string scale $M_s$ close to the  1-10 TeV range,
one interesting feature of the intersecting brane models is
the presence of a very well defined and model-independent class of 
extra TeV-scale $Z$' bosons.
Indeed, in all models there are some extra $U(1)$ symmetries beyond
hypercharge which seem rather model independent \cite{giiq}. 
We have generators $U(1)_a$ and $U(1)_d$ which are gauged 
baryon and lepton numbers respectively. $U(1)_c$ correspond 
to the 3$^{rd}$ component of right-handed weak isospin and $U(1)_b$ 
is a PQ-like gauged symmetry. In the  class of models 
in eq.(\ref{intersecMSSM}) the latter $U(1)_b$ symmetry is absent.
Hypercharge is a linear combination of $(B-L)$ and $U(1)_c$.
The orthogonal  $U(1)$'s may 
get Stueckelberg masses 
\footnote{ Note that all anomalous $U(1)$'s get masses through
this mechanism but also some anomaly-free $U(1)$'s may get
a mass, see ref.\cite{imr}.}
by combining with RR string fields $B_i^{\mu\nu}$,
as shown in fig. \ref{stuck}. In this way one gets a mass
matrix for the Abelian gauge bosons of the form:
\beq
M^2_{\alpha \beta } \ =\ \sum_{i} g_{\alpha} g_{\beta}
c_i^{\alpha }c_i^{\beta}M^2_s 
\eeq
where $\alpha , \beta $ run over the $U(1)$ factors of each specific model
and the  $c_i^{\alpha }$ are model-dependent 
coefficients which may be computed in each particular brane setting
\cite{giiq,elias}.
Thus for example,
in the class of models discussed in ref.\cite{imr} or
ref.\cite{cim3} one sees that this matrix has four
eigenvalues $M=(0,M_2,M_3,M_4)$, with the zero mode corresponding to
standard hypercharge. It turns out that in those models typically 
one of the eigenvalues is well below the string scale $M_s$, so
that one could detect the effects of such extra $U(1)$ 
before actually reaching the string threshold.  One can also 
put constraints on those eigenvalues from $\rho $-parameter
bounds \cite{giiq}.  It would be rather amusing if the first signature
of string physics would come from the detection of 
any of these particular extra Z' bosons.

\end{document}